\documentclass[onecolumn,showpacs,amsfonts,aps,prc,nofootinbib,floatfix,%
superscriptaddress]{revtex4}

\usepackage{amsmath}
\usepackage{bm}
\usepackage{graphicx}

\usepackage{epsfig}
\newcommand{\beq}{\begin{equation}}
\newcommand{\eeq}{\end{equation}}
\newcommand{\bea}{\vspace{0.25cm}\begin{eqnarray}}
\newcommand{\eea}{\end{eqnarray}}

\newcommand{\ro}{\mbox{{\boldmath
$\rho$}}}
\newcommand{\rr}{\mbox{{\boldmath
$\rho$}}}
\newcommand{\Pb}{\mbox{{\bf
P}}}

\newcommand{\nb}{\mbox{{\bf
n}}}
\newcommand{\bb}{\mbox{{\bf
b}}}

\newcommand{\pb}{\mbox{{\bf
p}}}
\newcommand{\rb}{\mbox{{\bf
r}}}
\newcommand{\kb}{\mbox{{\bf
k}}}

\newcommand{\qbt}{\mbox{{\bf
q}}_\perp}

\newcommand{\Bb}{\mbox{{\bf
B}}}

\newcommand{\Eb}{\mbox{{\bf
E}}}

\newcommand{\Fb}{\mbox{{\bf
F}}}

\newcommand{\fb}{\mbox{{\bf
f}}}
\newcommand{\vb}{\mbox{{\bf
v}}}

\def\lsim{\mathrel{\rlap{\lower4pt\hbox{\hskip1pt$\sim$}}
    \raise1pt\hbox{$<$}}}         
\def\gsim{\mathrel{\rlap{\lower4pt\hbox{\hskip1pt$\sim$}}
    \raise1pt\hbox{$>$}}}         

\newcommand{\landau}{L.D.~Landau Institute for Theoretical Physics,
        GSP-1, 117940, Kosygina Str. 2, 117334 Moscow, Russia}

\begin{document}


\title{
Effect of magnetic field on the photon radiation 
from quark-gluon plasma
in heavy ion collisions
}
\date{\today}

\author{B.G.~Zakharov}\affiliation{\landau}

\begin{abstract}
We develop a formalism for the photon emission
from the quark-gluon plasma with an external electromagnetic field.
We then use it to investigate the effect of magnetic field 
on the photon emission from the quark-gluon plasma created in $AA$ collisions.
We find that even for very optimistic assumption
on the magnitude of the magnetic field generated in $AA$ collisions
its effect on the photon emission rate is practically negligible.
For this reason the magnetic field cannot generate a significant 
azimuthal asymmetry in the photon spectrum.
\end{abstract}
%
\maketitle

\section{Introduction}
There is now a variety of experimental data on hadronic observables
in $AA$ collisions at RHIC and LHC that show that hadron production
in high energy $AA$ collisions goes via formation of a hot quark-gluon 
plasma (QGP) fireball. 
The major arguments in favor of the QGP formation at RHIC and LHC
are  the observation of a strong suppression of high-$p_{T}$ particle 
spectra (the so-called jet quenching phenomenon)
and the success of the hydrodynamical models in describing the
flow effects in hadron production in $AA$ collisions. 
The results of the jet quenching 
\cite{UW_JQMC,RAA12,JETC1,CUJET} 
and hydrodynamical \cite{Heinz_hydro1}  analyses support the production time
of the QGP $\tau_0\sim 0.5-1$ fm. However, this is only a 
qualitative estimate, because the value of $\tau_0$ is not well
constrained by the data on the jet quenching and the flow effects.
For jet quenching it is due
to a strong reduction of the radiative parton energy loss in
the initial stage of the QGP evolution by the finite size effects
\cite{Z_OA,Gale-Caron}. 
For this reason jet quenching is not very sensitive to the first
fm/c of the matter evolution.
And for the flow effects it is due to 
the low transverse velocities in the initial stage 
of the fireball evolution and the correlations of $\tau_0$ with 
the viscosity of the QGP in the hydrodynamical fits 
\cite{Heinz_hydro2,Heinz_tau}.

It is believed that the photon spectrum in the low and 
intermediate $k_T$ region may be more sensitive to the initial stage of
the QGP evolution than the hadronic observables.
Because the thermal photons radiated from the QGP leave the fireball 
without attenuation and the photon emission rate is largest in the 
initial hottest stage of the QGP evolution \cite{Shuryak}.
The measurements of the photon  
spectrum in $AA$ collisions
performed at RHIC 
\cite{PHENIX_ph1,PHENIX_ph_v2,PHENIX_ph_PR} 
and LHC \cite{ALICE_ph}
show that there is some excess of the photon
yield (above the photons from hadron decays and from the hard perturbative
mechanism)  at $k_T\lsim 3-4$ GeV.
It is widely believed
that it is related to the photon emission
from the QGP. However, the results of pQCD calculations
of the thermal contribution to the photon spectrum are only in a qualitative 
agreement with the data obtained at RHIC and LHC
(see \cite{Shen1} and references therein). Say, the theoretical predictions
obtained in recent analysis \cite{Gale_best} using a 
sophisticated viscous hydrodynamical model of the fireball evolution
underestimate the photon spectrum by a factor of $\sim 1.5-3.5$.
It was observed that the thermal photons exhibit a significant 
azimuthal asymmetry $v_2$ (elliptic flow)
comparable to that for hadrons.
It is difficult to reconcile this fact with the expectation that 
the thermal photons should be mostly radiated from the hottest initial stage
of the QGP where the flow effects should be small (this is often called
the direct photon puzzle).
It was suggested \cite{Dusling} 
that in the standard pQCD scenario of the thermal photon
emission the flow effect for photons may be related to 
the viscous effects in the QGP that lead to a deviation
of the parton distribution functions in the QGP from the equilibrium ones.
The numerical results of \cite{Gale_best} show that the viscosity of 
the QGP may be an important source of the photon momentum anisotropy.
However, in the analysis \cite{Gale_best} the viscous effects
have been accounted for only for the LO pQCD
$2\to 2$ processes $q(\bar{q})g\to \gamma q(\bar{q})$ (Compton)
and $q\bar{q}\to \gamma g$ (annihilation), 
and have not been included for the higher order collinear processes 
$q\to \gamma q$ and $q\bar{q}\to \gamma$ \cite{AMY1}.

The direct photon puzzle stimulated searches for novel mechanisms
of the photon production in $AA$ collisions that could generate
a significant azimuthal asymmetry.
In Ref. \cite{Kharzeev1} it was suggested that 
the large photon azimuthal anisotropy may be related
to a novel photon production mechanism stemming from the conformal
anomaly and a strong magnetic field in noncentral $AA$ collisions.
However, the contribution of this mechanism becomes important
only for a sufficiently large magnitude of the magnetic field, which
is not supported by calculations for realistic evolution of the plasma
fireball \cite{Z_maxw}.
In Ref. \cite{Snigirev} it was argued that the observed photon asymmetry
may be due to an intensive bremsstrahlung like synchrotron radiation
resulting from the interaction of escaping quarks with the collective
confining color field at the surface of the QGP. 
For this mechanism the asymmetry arises due to bigger surface 
emission from the almond-shaped QGP fireball along the direction
of the impact parameter vector (as shown in Fig.~1).
In Ref. \cite{T1} it was suggested that the significant photon $v_2$
can be related to the real synchrotron emission from the thermal quarks 
in a strong magnetic field generated in noncentral $AA$ collisions.
Since  the magnetic field in the noncentral $AA$ collisions is mostly
perpendicular to the reaction plane (this direction
corresponds to $y$ axis, if $x$ axis is directed along the impact parameter 
of the $AA$ collision as shown in Fig.~1) the synchrotron radiation 
rate is largest
in the direction along of the impact parameter vector. 
\begin{figure} [t]
\vspace{.7cm}
\begin{center}
\epsfig{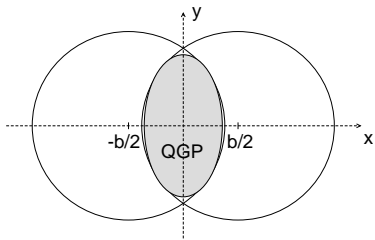}
\end{center}
\vspace{-0.5cm}
\caption[.]
{The transverse plane of a noncentral $AA$-collision with the impact
parameter $b$. 
}
\end{figure}
For this reason the synchrotron  mechanism  
leads naturally to a strong
azimuthal asymmetry of the photon emission.
This explanation works only if the contribution of 
the synchrotron mechanism to the photon emission rate is significant.
The analysis of Ref. \cite{T1} shows that 
in the central rapidity region at $k_T\sim 1-3$ GeV
the contribution of the synchrotron mechanism may be comparable
with the observed photon yield in Au+Au collisions at $\sqrt{s}=0.2$ TeV.
However, the calculations of Ref. \cite{T1} are of a qualitative nature.
In \cite{T1} 
the calculations are performed for purely synchrotron radiation.
But in the QGP each quark undergoes 
multiple scattering due to interaction with other thermal 
quarks and gluons. One can expect that it will lead to a 
reduction of the coherence/formation length of the photon
emission, and to suppression of the synchrotron emission.
In reality for the QGP with magnetic field one simply cannot 
distinguish between the synchrotron radiation and
the bremsstrahlung due to multiple scattering, and one has 
to treat both the mechanisms on an even footing. 
In this case the effect of magnetic field on the photon emission
can only be defined as  
the difference between the photon emission rate from the QGP 
with and without magnetic field. 
Also, in \cite{T1} the comparison with the experimental photon spectrum 
has been performed by integrating over the QGP four volume
neglecting the longitudinal and transverse 
expansion of the QGP.
The neglect of the longitudinal expansion of the QGP may be
too crude approximation. For a QGP with zero velocity the energy of a quark
radiating a photon with a given momentum is smaller 
than that in the comoving frame for the QGP with the longitudinal expansion. 
Since the quark (anti-quark) thermal distribution
decreases exponentially with quark energy, the approximation
of zero QGP velocity can overestimate considerably the photon spectrum.
Another issue that can result in overestimation of the synchrotron contribution
is the use in \cite{T1} of the current quark
masses. In the QGP quarks acquire a thermal quasiparticle mass $\sim gT$,
that appears after the Hard Thermal Loop (HTL) resummation
(which is very important already for the LO $2\to 2$ processes 
\cite{Baier_ph}). 
Since the synchrotron spectrum reduces with the charged particle mass,
the accounting of the quark quasiparticle mass, that is much bigger
than the current quark masses, should suppress considerably
the effect of magnetic field. 

Besides the photon bremsstrahlung addressed in \cite{T1} 
the magnetic field can affect the photon production via
the annihilation  mechanism $q\bar{q}\to \gamma$.
The analysis of the collinear processes $q\to \gamma q$ and $q\bar{q}\to
\gamma$ for the QGP without magnetic field 
shows that 
the annihilation contribution is even more important than bremsstrahlung 
at the photon momenta $k \gg T$ \cite{AMY1}. 
%
The purpose of the present work is to address the effect of magnetic
field on both the processes $q\to \gamma q$ and $q\bar{q}\to \gamma$
(below we will call the magnetic field modification for both these processes
as the synchrotron contribution).
We develop a formalism which treats on an even footing 
the effect of multiple scattering and curvature of the quark trajectories
in the collective magnetic field in the QGP. 
Our analysis is based on the light cone
path integral (LCPI) formalism \cite{LCPI}, which was previously
successfully used \cite{AZ} for a very simple derivation
of the photon emission rate from the higher order collinear processes
$q\to \gamma q$ and $q\bar{q}\to \gamma$ obtained earlier 
by Arnold, Moore and Yaffe
(AMY) \cite{AMY1} using methods from thermal field theory 
with the HTL resummation.
It is known that the higher order diagrams 
corresponding to these processes contribute to leading order \cite{AGZ2000},
and turn out to be as important as the LO  
$2\to 2$ processes $q(\bar{q})g\to \gamma q(\bar{q})$
and $q\bar{q}\to \gamma g$.
Contrary to the collinear processes the LO processes 
should not be affected by the presence of an external magnetic field.
Our results differ drastically from that of 
\cite{T1}. We find that even for very optimistic
magnitude of the magnetic field for RHIC and LHC conditions
its effect on the photon
emission from the QGP is very small.

The plan of the paper is as follows. In Sec.~2 we first 
discuss the physical picture of the processes
$q\to \gamma q$ and $q\bar{q}\to \gamma$. We show
that for the magnitude of the magnetic field of interest
for $AA$ collisions these process remain in the collinear
regime. 
Then we develop a formalism for evaluation of 
their contribution to the photon emission from the QGP
with magnetic field in the medium rest frame.
In Sec.~3 we discuss how to compute the photon spectrum
from the plasma fireball in $AA$ collisions.
We discuss the model of the fireball and the possible
magnitude of the magnetic field for the most optimistic
scenario for the synchrotron photon emission.
In Sec.~4  we present our numerical results.
Sec.~5 summarizes our work.
Some of our results concerning the photon emission
rate from the QGP at rest have been reported
in an earlier short communication \cite{Z_JETPL}.

\section{Bremsstrahlung and pair annihilation 
in the QGP with magnetic field}
In this section we discuss the photon emission rate per unit time and volume
in the equilibrium QGP with magnetic field in the QGP rest frame. 
Similarly to the analyses \cite{AMY1,AZ} of the 
processes $q\to \gamma q$ and $q\bar{q}\to \gamma$
for zero magnetic field we treat quarks and photons as relativistic 
quasiparticles with energies much larger than their quasiparticle 
masses $m_{q}$ and $m_{\gamma}$ \footnote{We assume that the photons 
emitted in the QGP  adiabatically become massless after escaping from
the plasma fireball.}.  
For the weakly coupled QGP with $N_f$ flavors 
$m_q$ and $m_{\gamma}$ read \cite{AMY1}
\beq
m_q=gT/\sqrt{3}\,,
\label{eq:10}
\eeq 
\beq
m_{\gamma}=\frac{eT}{3}\sqrt{(3+N_f)/2}\,,
\label{eq:20}
\eeq 
where $g=\sqrt{4\pi\alpha_s}$ is the QCD coupling constant,
$e$ is the electron charge. In numerical calculations
we take $N_f=2.5$ to account for qualitatively the
suppression of strange quarks at moderate temperatures.
Since $m_q/m_{\gamma}>>1$ the effect of the nonzero photon mass
is very small, and our results are close to that for massless photon.

\subsection{Physical picture of photon emission and photon formation length}
The physical picture behind the derivation of the photon emission
rate in the QGP without magnetic field 
from the processes $q\to \gamma q$ and $q\bar{q}\to \gamma$
given in \cite{AMY1,AZ} is the fact that
in the weakly coupled QGP the hard partons with energy $E\gsim T$ 
undergo typically only small angle multiple 
scattering due to interaction with 
the random  soft gluon fields at the momentum scale $\sim gT$. 
And the large angle scattering with the momentum transfer $\sim E$ 
is a very rare process.
The typical quark scattering angle at the longitudinal scale
about the photon coherence/formation length, $L_f$, is small \cite{AZ}.
Due to this fact the processes $q\to \gamma q$ and 
$q\bar{q}\to \gamma$ are dominated 
by the collinear configurations, when the photon is emitted practically 
in the direction of the initial quark for $q\to \gamma q$ (and
in the direction of the momentum of the $q\bar{q}$ pair
for $q\bar{q}\to \gamma$). 
For a QGP with magnetic field this picture will remain valid if
\beq
\frac{L_f}{R_L}\ll 1\,,
\label{eq:30}
\eeq
where $R_L=E_q/z_qeB$ is the quark Larmor radius in
the magnetic field ($z_q$ is the quark electric charge in units of $e$). 
Let us demonstrate that the condition (\ref{eq:30})
is satisfied for the fields $eB=cm_{\pi}^2$ with $c\lsim 1$
that are of interest for $AA$ collisions.
Making use the formulas of the LCPI approach for the 
bremsstrahlung due to multiple scattering \cite{LCPI} and 
for the synchrotron emission \cite{Z_sync_QCD} 
one can obtain qualitative estimate 
\beq
L_f\sim \text{min}(L_1,L_2)\,,
\label{eq:40}
\eeq
where the quantities $L_{1,2}$ 
read
\beq
L_1\sim\frac{2E_q(1-x)S_{LPM}}{m_q^2x}\,,
\label{eq:50}
\eeq
\beq
L_2\sim\left(\frac{24E_qx(1-x)}{f^2}\right)^{1/3}\,.
\label{eq:60}
\eeq
Here $S_{LPM}$ is the suppression factor due to the 
Landau-Pomeranchuk-Migdal (LPM) effect \cite{LP,Migdal}, $x$ is the photon fractional
longitudinal momentum,  
$f=z_qx eB$.
For $S_{LPM}=1$ (\ref{eq:40}) gives simply the formation length
for the synchrotron emission in vacuum \cite{Z_sync_QCD}.
The LPM suppression factor can be easily estimated in the 
oscillator approximation corresponding to the 
description of multiple scattering in terms of
the transport coefficient $\hat{q}$ in the BDMPS \cite{BDMPS}
approach to the induced gluon emission.
In the oscillator approximation 
$S_{LPM}\sim \frac{3}{\kappa \sqrt{2}}$ (see below (\ref{eq:500})) \cite{LCPI},
where $\kappa=[8\hat{q}E_q (1-x)/9xm_q^4]^{1/2}$ (we take here $m_\gamma=0$)
A qualitative pQCD estimate gives $\hat{q}\sim 14T^3$ (see below).
From the point of view of the photon emission from the QGP
the interesting $x$-region is $x\gsim 0.5$.
Making use of (\ref{eq:50}), (\ref{eq:60}) 
one can obtain for $eB=c m_{\pi}^2$ 
at $x\sim 0.5$ for $u$ quark ($z_q=2/3$)
\beq
L_1\sim \frac{1}{T}\sqrt{E_q/T}\,,
\label{eq:70}
\eeq
\beq
L_2\sim 4\left(\frac{E_q}{c^2m_{\pi}^4}\right)^{1/3}\,.
\label{eq:80}
\eeq
From (\ref{eq:70}) and (\ref{eq:80}) one can see that for the QGP
temperatures $T\gsim T_c$
(here $T_c\approx 160-170$ MeV is the deconfinement temperature \cite{EoS})
we have $L_1<L_2$ in the energy region of interest $E_q\lsim 5$ GeV, i.e.
we have $L_f\sim L_1$. Then we obtain
\beq
\frac{L_f}{R_L}\sim  z_q c\left(\frac{m_{\pi}}{T}\right)^{3/2}
\left(\frac{m_{\pi}}{E_q}\right)^{1/2}
\label{eq:81}
\eeq
From (\ref{eq:81}) for $c=1$ we obtain that 
at $E_q\gsim 1$ GeV for $u$ quark $L_f/R_L\lsim 0.25 (m_{\pi}/T)^{3/2}$.
Thus the condition (\ref{eq:30}) is reasonably satisfied even at $T\sim T_c$. 
The contribution
of the annihilation $q\bar{q}\to \gamma$ may be expressed
via the spectrum of the $\gamma\to q\bar{q}$ transition (see below).
By repeating the above estimates for $\gamma\to q\bar{q}$
one can show that for this case the condition (\ref{eq:30}) is also
satisfied.

\subsection{Basic formulas}
The above analysis shows that, similarly to the QGP without magnetic
field \cite{AMY1,AZ},   
for the QGP produced in $AA$ collisions
in the presence of magnetic field
we can treat 
the processes $q\to \gamma q$ and $q\bar{q}\to \gamma$ 
as the collinear ones. 
And the contribution of these processes to the photon emission 
rate per unit time and volume in the plasma rest frame 
can be written as \cite{AZ,AMY1}
\beq
\frac{dN}{dtdVd\kb}=
\frac{dN_{br}}{dtdVd\kb}
+\frac{dN_{an}}{dtdVd\kb}\,,
\label{eq:90}
\eeq
where the first term corresponds to $q\to \gamma q$ and the second one
to $q\bar{q}\to \gamma$.
The bremsstrahlung contribution reads \cite{AZ}
\bea
\frac{dN_{br}}{dtdVd\kb}=\frac{d_{br}}{k^{2}(2\pi)^{3}}
\sum_{s}
\int_{0}^{\infty} dp p^{2}n_{F}(p)
[1-n_{F}(p-k)]\theta(p-k)
\frac{dP^{s}_{q\rightarrow \gamma q}(\pb,\kb)}{dk dL}\,,
\label{eq:100}
\eea
where 
$d_{br}=4N_{c}$ is the number of the quark and antiquark states,
\beq
n_{F}(p)=\frac{1}{\exp(p/T)+1}\,
\label{eq:110}
\eeq
is the thermal Fermi distribution, 
and 
${dP^{s}_{q\rightarrow \gamma q}(\pb,\kb)}/{dk dL}$
is the probability distribution of the photon emission 
in the QGP per unit length from a fast quark of type $s$.
Since we work in the small angle approximation, 
we can take the vectors $\pb$  and $\kb$ parallel.
The quantity $dP^{s}_{q\rightarrow \gamma q}(\pb,\kb)/{dk dL}$
should be evaluated accounting for the quark interaction with the 
random soft gluon field generated by the thermal partons and 
with the smooth external electromagnetic field. 

The annihilation contribution 
can be expressed via the probability
distribution for the photon absorption 
$dN_{abs}/dtdVd\kb$ with
the help of the detailed balance principle which gives 
\cite{AZ}
\beq
\frac{dN_{an}}{dtdVd\kb}=
[1+n_{B}(k)]^{-1}
\frac{dN_{abs}}{dtdVd\kb}\,,
\label{eq:120}
\eeq
where $n_{B}(k)=1/[\exp(k/T)-1]$ is the Bose distribution.
The photon absorption rate on the right-hand side of (\ref{eq:120})
can be written as
\bea
\frac{dN_{abs}}{dtdVd\kb}=\frac{d_{an}n_{B}(k) }{(2\pi)^{3}}
\sum_{s}
\int_{0}^{\infty} dp [1-n_{F}(p)]
[1-n_{F}(k-p)]\theta(k-p)
\frac{dP^{s}_{\gamma\rightarrow q\bar{q}}(\kb,\pb)}{dp dL}\,,
\label{eq:130}
\eea
where $d_{an}=2$ is the number
of the photon helicities, 
$
{dP^{s}_{\gamma\rightarrow q\bar{q}}(\kb,\pb)}/{dp dL}
$ is the probability distribution per unit length
for the $\gamma \rightarrow q\bar{q}$ transition
($p$ is the quark momentum and $k-p$ is the antiquark momentum,
and similarly to $q\to \gamma q$ we can take the vectors 
$\pb$  and $\kb$ parallel).
Using the relation
\bea
\frac{n_{B}(k)}{1+n_B(k)}[1-n_{F}(p)]
[1-n_{F}(k-p)]
=n_{F}(p)
n_{F}(k-p)\,
\label{eq:140}
\eea
from (\ref{eq:120}), (\ref{eq:130}) one obtains \cite{AZ}
\bea
\frac{dN_{an}}{dtdVd\kb}=\frac{d_{an} }{(2\pi)^{3}}
\sum_{s}
\int_{0}^{\infty} dp n_{F}(p)
n_{F}(k-p)\theta(k-p)
\frac{dP^{s}_{\gamma\rightarrow q\bar{q}}(\kb,\pb)}{dp dL}\,.
\label{eq:150}
\eea

Let us consider first calculation of the bremsstrahlung contribution. 
In the LCPI formalism \cite{LCPI} 
the probability of the $q\to \gamma q$ transition
(for a quark with charge $z_{q}e$) per unit length can be written
in the form
(we use here the fractional photon momentum $x$ instead of 
$k$)
\bea
\frac{d P_{q\rightarrow \gamma q}}{d
x dL}=2\mbox{Re}
\int\limits_{0}^{\infty} d
z
\exp{\left(-i\frac{z}{\lambda_{f}}\right)}
\hat{g}(x)\left[
{\cal K}(\ro_{2},z|\ro_{1},0)
-{\cal K}_{vac}(\ro_{2},z|\ro_{1},0)
\right]\bigg|_{\ro_{1,2}=0}\,,
\label{eq:160}
\eea
where $\lambda_{f}=2M(x)/\epsilon^{2}$ with $M(x)=E_qx(1-x)$,
$\epsilon^{2}=m_{q}^{2}x^{2}+m_{\gamma}^{2}(1-x)$
(in general for $a\to b+c$ transition 
$\epsilon^{2}=m_{b}^{2}x_{c}+m_{c}^{2}x_{b}-m_{a}^{2}x_{b}x_{c}$),
$\hat{g}$ is the vertex operator, given by
\beq
\hat{g}(x)=\frac{V(x)}{M^{2}(x)}\frac{\partial }{\partial \ro_{1}}\cdot
\frac{\partial }{\partial \ro_{2}}\,
\label{eq:170}
\eeq
with 
\beq
V(x)=z_{q}^{2}\alpha_{em}(1-x+x^{2}/2)/x, 
\label{eq:180}
\eeq
$\alpha_{em}=e^2/4\pi$ the fine-structure constant. $\cal{K}$ in (\ref{eq:160}) is the retarded Green function of a two dimensional
Schr\"odinger equation, in which the longitudinal coordinates $z$ 
(along the initial quark momentum) plays the role of time, 
with the Hamiltonian
\beq
\hat{\cal{H}}=-\frac{1}{2M(x)}
\left(\frac{\partial}{\partial \ro}\right)^{2}
+         v(\ro)\,,
\label{eq:190}
\eeq
and 
\beq
{\cal{K}}_{vac}(\ro_{2},z|\ro_{1},0)=\frac{M(x)}{2\pi iz}
\exp\left[\frac{iM(x)(\ro_2-\ro_1)^2}{2z}\right]
\label{eq:200}
\eeq
is the Green function for $v=0$.
The potential $v$ can be written as
\beq
v=v_{f}+v_{m}\,,
\label{eq:210}
\eeq
where $v_{f}$ is due to the fluctuating gluon fields of the QGP,
and $v_{m}$ is related to the mean electromagnetic field.
The mean field component of the potential reads
\beq
v_{m}=-\fb \ro\,,
\label{eq:220}
\eeq
where $\fb=x z_{q}\Fb$, $\Fb$ is transverse component (to the parton momentum) 
of the Lorentz force for a particle with charge $e$.
The effect of the longitudinal Lorentz force (which exists for nonzero
electric field) is small for the relativistic partons,
and we neglect it.
The component $v_f$ reads  
\beq
v_{f}=-i P(x\rho)\,.
\label{eq:230}
\eeq
Here the function $P(\rho)$ 
can be written as
\beq
P(\ro)=g^{2}C_{F}\int\limits_{-\infty}^{\infty} dz 
[G(z,0_{\perp}z)-G(z,\ro,z)]\,,
\label{eq:240}
\eeq
where $g$ is the QCD coupling, $C_F=4/3$ is the quark Casimir,
$G$ is the gluon correlator (the color indexes are omitted)
\beq
G(x-y)= 
u_{\mu}u_{\nu}
{\Large\langle\Large\langle}
A^{\mu}(x)A^{\nu}(y)
{\Large\rangle\Large\rangle}\,.
\label{eq:250}
\eeq
Here $u^{\mu}=(1,0_{\perp},1)$ is the light-like four vector along
the $z$ axis.
The gluon correlator $G$ may be expressed via the HTL gluon polarization
operator. Making use of an elegant sum rule for the transverse and longitudinal
HTL gluon self-energies derived in \cite{PA_C}
the function $P(\ro)$ may be written as \cite{AZ}  
\beq
P(\ro)= \frac{g^{2}C_{F}T}{(2\pi)^{2}}\int d\qbt [1-\exp(i\ro \qbt)]
C(\qbt)\,,
\label{eq:260}
\eeq
\beq
C(\qbt)=\frac{m_{D}^{2}}{\qbt^{2}(\qbt^{2}+m_{D}^{2})}\,,
\label{eq:270}
\eeq  
where $m_{D}=gT[(N_{c}+N_{F}/2)/3]^{1/2}$ 
is the Debye mass. 
In \cite{AZ} it was demonstrated that for the case without external field 
calculation of the spectrum given 
by (\ref{eq:160}) within the LCPI formalism with 
the use of (\ref{eq:260}), (\ref{eq:270}) is equivalent to solving the integral
equation obtained in the AMY analysis \cite{AMY1} 
in the momentum representation. And the formulas (\ref{eq:90}),
(\ref{eq:100}), (\ref{eq:150}) reproduce exactly the AMY photon 
emission rate.

In the approximation of static color Debye-screened scattering centers 
(in the sense of quark multiple scattering in the QGP)
\cite{GW} the function $P(\ro)$ reads
\beq
P(\ro)= 
\frac{n{\sigma}_{q\bar{q}}(\rho )}{2}\,,
\label{eq:280}
\eeq
where $n$ is the number density
of the color centers, and 
\beq
\sigma_{q\bar{q}}(\rho)={C_{T}C_{F}\alpha_{s}^{2}}\int d\qbt
\frac{[1-\exp(i\qbt\ro)]}{(\qbt^{2}+m_{D}^{2})^{2}}\,\,
\label{eq:290}
\eeq
is the well known dipole cross section \cite{NZ12}
with $C_T$ being the color center Casimir.

Both for the HTL scheme (\ref{eq:260}), (\ref{eq:270}) and the 
static approximation (\ref{eq:280}), (\ref{eq:290}) at $\rho\lsim 1/m_D$ 
approximately $P(\rho)\propto \rho^2$. At $\rho\ll 1/m_D$ the function
$P(\rho)$ in the static model differs from that in the HTL scheme
just by the normalization factor 
$\frac{\pi^{2}}{6\cdot 1.202}(1+N_{f}/6)/(1+N_{f}/4)
\approx 1.19$ (for $N_{f}=2.5$).
The replacement of the factor $1/(q^{2}+m_{D}^2)^{2}$ in the dipole 
cross section in the static model
by $1/[q^{2}(q^{2}+m_{D}^2)]$ in the HTL scheme
leads to unlimited growth of $P(\rho)$ at large
$\rho$ (due to zero magnetic mass in the HTL approximation), 
while for static model $P(\rho)$ flattens at $\rho\gsim 1/m_{D}$.
However, this difference is not very important from the point
of view of the photon emission, because the contribution
of the region $\rho\gsim 1/m_D$ is relatively small (in the sense
of the path integral representation of the Green function $\cal{K}$
entering to (\ref{eq:160})). 

We will work in the oscillator approximation 
\beq
P(\rho)=C_{p}\rho^{2}\,,
\label{eq:300}
\eeq
which is widely used in jet quenching analyses 
\cite{JQ_OA1,JQ_OA2,JQ_OA3,JQ_OA4,JQ_OA5,JQ_OA6}.
The $C_{p}$ can be expressed via  the transport 
coefficient $\hat{q}$ \cite{BDMPS}, describing 
gluon transverse momentum broadening in the QGP, as 
$C_p=\hat{q}C_{F}/4C_{A}$.
In numerical calculations we use $\hat{q}\propto T^3$ and set
$\hat{q}=0.2$ GeV$^3$ at $T=250$ MeV. 
This value is supported by estimate of $\hat{q}$ 
within the static model via the magnitude of 
the dipole cross section at $\rho\sim 1/m_q$ 
that allows to describe well the data on jet quenching 
in $AA$ collisions within the LCPI scheme \cite{RAA12}.
It also agrees with the qualitative pQCD calculations of 
Ref.~\cite{Baier_qhat} that give
$
\hat{q}\sim 2\varepsilon^{3/4}\,,
$
with $\varepsilon$ the QGP energy density) (it gives 
$\hat{q}\approx 14T^3$). 
Note that the estimate obtained in \cite{Baier_qhat} agrees 
with the relation between $\hat{q}$ and the ratio of the 
shear viscosity to the entropy density
 $\eta/s$
\beq
\hat{q}\sim 1.25 T^3 s/\eta
\label{eq:310}
\eeq 
obtained in \cite{MMW_qhat} if one takes the quantum limit value
$\eta/s=1/4\pi$ \cite{Son}.

\subsection{Photon spectrum in the oscillator approximation}
For the quadratic $P(\rho)$  the Hamiltonian (\ref{eq:190})
takes the oscillator form (we omit arguments of functions for brevity,
where possible)
\beq
\hat{\cal{H}}=-\frac{1}{2M}
\left(\frac{\partial}{\partial \ro}\right)^{2}
+   \frac{M\Omega^{2}\ro^2}{2}-\fb\ro\,
\label{eq:320}
\eeq
with
\beq
\Omega=\sqrt{-iC_{p}x^{2}/M}\,.
\label{eq:330}
\eeq
The Green function for the Hamiltonian (\ref{eq:320}) is known explicitly 
(see, for example, \cite{FH})
\beq
{\cal{K}} (\rr_2 , z_2 | \rr_1 , z_1)=
\frac{M\Omega}{2\pi i\sin(\Omega  z)}
\exp{[i S_{cl}(\rr_2 , z_2 | \rr_1 , z_1)]}\,,
\label{eq:340}
\eeq
where $ z=z_2-z_1$, and $S_{cl}$ is the classical action. 
The action can be
written as a sum  $S_{cl}=S_{osc}+S_{f}$ with
\beq
\hspace{-.05cm}S_{osc}(\rr_2 , z_2 | \rr_1 , z_1)=
\frac{M\Omega}{2\sin(\Omega  z)}
\left[\cos(\Omega  z)(\rr_{1}^{2}+\rr_{2}^{2})-2\ro_{1}\ro_{2}\right]\,,
\label{eq:350}
\eeq
\beq
S_{f}(\rr_2 , z_2 | \rr_1 , z_1)
=
\frac{M\Omega}{2\sin(\Omega  z)}
\left[\Pb(\ro_{1}
+\ro_{2})-W\right]\,,
\label{eq:360}
\eeq
where
\beq
\Pb=\frac{2\fb\, [1-\cos(\Omega  z)] }{M\Omega^{2}}
\,\,,
\label{eq:370}
\eeq
\beq
W=\frac{2\fb^{2}}{M^{2}\Omega^{4}}
\left[1-\cos(\Omega  z)-\frac{\Omega z
\sin(\Omega  z)}{2}\right]\,.
\label{eq:380}
\eeq

Then, after including the vacuum term in (\ref{eq:160}), 
 a simple calculation gives
\beq
\frac{dP}{dxdL}=2V(x) (I_{osc}+\Delta I)\,.
\label{eq:390}
\eeq
Here $I_{osc}$ corresponds to the pure oscillator case
($\fb=0$). It reads
\bea
I_{osc}=\frac{1}{\pi}\mbox{Re}\int_{0}^{\infty} dz
\left[
\frac{1}{z^2}-\left(\frac{\Omega}{\sin(\Omega z)}\right)^{2}
\right]
\exp\left(-i\frac{z}{\lambda_{f}}\right)\,.
\label{eq:400}
\eea
And $\Delta I$ gives the synchrotron correction.
It can be written as a sum $\Delta I=I_1+I_2$ with 
\beq
I_{1}=
\frac{1}{\pi}\mbox{Re}\int_{0}^{\infty} dz
\left(\frac{\Omega}{\sin(\Omega z)}\right)^{2}
[1-\exp(-U)]\exp\left(-i\frac{z}{\lambda_{f}}\right)\,,
\label{eq:410}
\eeq
\bea
I_{2}=
\frac{1}{\pi}\mbox{Re}\int_{0}^{\infty} dz
\frac{iM\Omega^{3}}{8\sin^{3}(\Omega z)}
\Pb^2
\exp\left(-U-i\frac{z}{\lambda_{f}}\right)\,,
\label{eq:420}
\eea
where
\beq
U=\frac{iM\Omega W}{2\sin(\Omega z)}\,.
\label{eq:430}
\eeq

For numerical calculations it is convenient to introduce 
the dimensionless integrals
\beq
\bar{I}_{osc,1,2}=\frac{\pi}{|\Omega|}I_{osc,1,2}\,,
\label{320}
\eeq
and to use the dimensionless integration variable $\tau=z|\Omega|\exp(i\pi/4)$.
Then we obtain for $\bar{I}_{osc,1,2}$ 
\beq
\bar{I}_{osc}(\kappa)=
\mbox{Re}
\int_{0}^{\infty} \frac{d\tau\exp(i\pi/4)}{\tau^{2}}
\left(1-\frac{\tau^{2}}{\sinh^{2}\tau}\right)
\exp\left(-\frac{(1+i)\tau}{\sqrt{2}\kappa}\right)\,,
\label{eq:450}
\eeq
\beq
\bar{I}_{1}(\kappa,\phi)=
\mbox{Re}
\int_{0}^{\infty} \frac{d\tau\exp(i\pi/4)}{\sinh^{2}\tau}
\left[1-\exp(-U)\right]
\exp\left(-\frac{(1+i)\tau}{\sqrt{2}\kappa}\right)\,,
\label{eq:460}
\eeq
\beq
\bar{I}_{2}(\kappa,\phi)=
\frac{\phi}{2}
\mbox{Re}
\int_{0}^{\infty}d\tau
 \frac{(1-\cosh \tau)^{2}}{\sinh^{3}\tau}
\exp\left(-\frac{(1+i)\tau}{\sqrt{2}\kappa}-U\right)\,,
\label{eq:470}
\eeq
where now
\beq
U=\frac{(1-i)\phi}{2\sqrt{2}}
\left[\tau-2\tanh(\tau/2)\right]\,,
\label{eq:480}
\eeq
and the dimensionless parameters $\kappa$ and $\phi$ read  
$\kappa=\lambda_{f}|\Omega|$, $\phi=\fb^{2}/M|\Omega|^{3}$.

In the low density limit ($\kappa\to 0$)
$\bar{I}_{osc}(\kappa)\approx \kappa/3$.
The higher order terms in $\kappa$ describe the LPM effect.
The ratio of $\bar{I}_{osc}$ to the leading order term gives the 
LPM suppression factor 
\beq
S_{LPM}=3\bar{I}_{osc}/\kappa\,.
\label{eq:490}
\eeq
From (\ref{eq:450}), (\ref{eq:490}) one can obtain for two 
limiting cases of strong ($\kappa\gg 1$)
and weak ($\kappa\ll 1$) LPM effect \cite{LCPI}:
\beq
S_{LPM}\approx \frac{3}{\kappa\sqrt{2}}\,\,\,(\kappa\gg 1)\,,\,\,\,\,
S_{LPM}\approx 1-\frac{16\kappa^4}{21}\,\,\,(\kappa\ll 1)\,.
\label{eq:500}
\eeq

In the limit $\Omega\to 0$ $I_{osc}=0$ and the integrals $I_{1,2}$
(\ref{eq:410}), (\ref{eq:420}) take the form (we denote them $I_{1,2}^s$)
\beq
I_{1}^s=
\frac{1}{\pi}\mbox{Re}\int_{0}^{\infty} \frac{dz}{z^2}
\exp\left(-i\frac{z}{\lambda_{f}}\right) 
\left[1
-\exp\left(-i\frac{\fb^{2}z^{3}}{24M}\right)
\right]\,,
\label{eq:510}
\eeq
\beq
I_{2}^s=
\frac{1}{\pi}\mbox{Re}\int_{0}^{\infty} dz
\frac{i\fb^{2}z}{8M}
\exp\left(-i\frac{z}{\lambda_{f}}-i\frac{\fb^{2}z^{3}}{24M}\right)
\,.
\label{eq:520}
\eeq
Similarly to the case of $I_{1,2}$ (\ref{eq:410}), (\ref{eq:420})   
it is convenient to go
from (\ref{eq:510}), (\ref{eq:520}) to 
dimensionless integrals. Now we define them as 
\beq
\bar{I}_{1,2}^s=\pi \lambda_{f} I_{1,2}^{s}\,.
\label{eq:530}
\eeq
Using the dimensionless integration variable
$\tau=z\exp(i\pi/4)/\lambda_{f}$ from (\ref{eq:410}), (\ref{eq:420}) 
taking the limit $\Omega\to 0$
we obtain 
\beq
\bar{I}_{1}^{s}(\phi_{s})=
\mbox{Re}
\int_{0}^{\infty} \frac{id\tau}{\tau}
\left[\exp{\left(-\frac{(1-i)\phi_{s}\tau^{3}}{\sqrt{2}}\right)}-1\right]
\exp\left(-\frac{(1+i)\tau}{\sqrt{2}}\right)\,,
\label{eq:540}
\eeq
\beq
\bar{I}_{2}^{s}(\phi_{s})=6\phi_{s}
\mbox{Re}
\int_{0}^{\infty} d\tau\tau
\exp\left(-\frac{(1+i)\tau}{\sqrt{2}}-
\frac{(1-i)\phi_{s}\tau^{3}}{\sqrt{2}}\right)\,,
\label{eq:550}
\eeq
where $\phi_{s}=\fb^{2}\lambda_{f}^{3}/24M\,$.
Functions (\ref{eq:540}), (\ref{eq:550}) may be expressed via
the Airy function 
$\mbox{Ai}(z)=\frac{1}{\pi}\sqrt{\frac{z}{3}}K_{1/3}(2z^{3/2}/3)$ (here
$K_{1/3}$ is the Bessel function) 
\beq
\bar{I}_{1}^{s}(\phi_s)=
-\pi\int_{z}^{\infty} dt \mbox{Ai}(t)\,,
\label{eq:560}
\eeq
\beq
\bar{I}_{2}^{s}(\phi_s)=
-\frac{2\pi}{z}\mbox{Ai}^{'}(z)\,,\,\,\,\,\,\,\,\,
\label{eq:570}
\eeq
where $z=1/(3\phi_s)^{1/3}$.
Our probability of photon emission in the  limit $\Omega\to 0$ 
is reduced to the well known quasiclassical formula
for the synchrotron spectrum \cite{BK,LL4} in QED.

For $\gamma\to q \bar{q}$ one can obtain similar formulas.
But now $M(x)=E_{\gamma}x(1-x)$ ($x$ is the quark fractional momentum)
$\epsilon^{2}=m_{q}^{2}-m_{\gamma}^{2}x(1-x)$,
$\fb=z_{q}\Fb$, and
\beq
V(x)=z_{q}^{2}\alpha_{em}N_c[x^{2}+(1-x)^{2}]/2\,,
\label{eq:580}
\eeq
\beq
\Omega=\sqrt{-iC_{p}/M}\,.
\label{eq:590}
\eeq
The factor $N_c$ in (\ref{eq:580}) accounts for summing
over the quark color indices for $\gamma \to q\bar{q}$ process.
For $q\to \gamma q$ it does not appear in (\ref{eq:180}) since
the sum over the quark color states is included in the 
factor $d_{br}$ in (\ref{eq:100}). 

Note that for the contribution of multiple scattering alone  the
oscillator approximation is equivalent to Migdal's calculations
in QED within the Fokker-Planck approximation \cite{Migdal}.
The oscillator approximation can lead to large errors 
in description of the gluon/photon emission from fast partons
produced in hard reactions in the regime when the formation length
is much bigger than the QGP size \cite{Z_OA,Z_phot}. 
In this regime the oscillator approximation underestimates strongly
the gluon/photon spectrum.
However, this problem does not arise for the photon
emission by the thermal quarks. In this case we have a situation 
similar to that for the photon emission from a quark propagating 
in an infinite medium.
In this regime the errors of the oscillator approximation should not
be large. 

\section{Photon spectrum in $AA$ collisions}
\subsection{Integration over space-time coordinates}
For the $AA$ collision at a given impact parameter $b$
the thermal contribution to the photon spectrum
$dN/dyd\kb_T$ (we will consider the central rapidity region $y=0$)
can be written 
as
\beq
\frac{dN}{dy d\kb_T}=\int dtdV \,\omega'\frac{dN(T',F',k')}{dt'dV'd\kb'}\,,
\label{eq:600}
\eeq
where primed quantities correspond to the comoving frame, and 
$\omega'=k'=|\kb'|$
(here we consider a photon as a massless particle). 
In (\ref{eq:600}) we write explicitly the arguments of the photon emission
rate in the comoving frame. The argument
$F'$ is the absolute value of the transverse (to the direction
of the emitted photon) Lorentz force acting on a particle with 
electric charge $e$.
Note that the photon emission rate in the comoving frame 
does not depend directly on the azimuthal direction 
of the photon momentum, and angular dependence of the left
hand side of (\ref{eq:600}) stems solely from   
the dependence of the photon emission rate $dN(T',F',k')/dt'dV'd\kb'$ 
on the right hand side
on the photon momentum $k'$ and on the Lorentz force
$F'$.
The value of $\omega'$ may be written via the photon four momentum
$k^{\mu}=(\omega,\kb_T,0)$ in the c.m. frame of the $AA$ collision as
\beq
\omega'=u^{\mu}k_{\mu}\,,
\label{eq:610}
\eeq
where 
\beq
u^{\mu}=\left(\frac{1}{\sqrt{1-\vb^2}},\frac{\vb}{\sqrt{1-\vb^2}}\right)
\label{eq:620}
\eeq
is the four velocity of the QGP cell. 
The value of $F'$ also can be expressed 
via the photon four momentum $k^{\mu}$ and the four velocity of the QGP cell.
In the matter comoving frame 
\beq
F'=e\left|\Eb_{\perp}^{'}+[\nb^{'}\times \Bb^{'}]\right|\,,
\label{eq:630}
\eeq
where $\nb^{'}$ is the unit vector in the direction of 
the photon momentum, $\Eb_\perp$ is tranverse (to the vector $\nb'$) 
component of the electric field. 
In terms of the electromagnetic field tensor
$F^{\mu\nu}$ in the c.m. frame of the $AA$ 
collisions (\ref{eq:630}) can be written
as
\beq
F'=e\sqrt{-L^{\mu}L_{\mu}}\,,
\label{eq:640}
\eeq
where
\beq
L^{\mu}=\frac{F^{\mu\nu}k_{\mu}}{u^{\delta}k_{\delta}}\,.
\label{eq:650}
\eeq
 
As usual we write the four volume integration in (\ref{eq:600}) changing
the integration variables $t$, $z$ to the  
proper time $\tau$ and rapidity $Y$ 
\beq
\tau=\sqrt{t^2-z^2}\,,\,\,\,\, 
Y=\frac{1}{2}\ln\left(\frac{t+z}{t-z}\right)\,.
\label{eq:660}
\eeq
In these coordinates 
\beq
\frac{dN}{dy d\kb_T}=\int \tau d\tau dY d\ro\, \omega'
\frac{dN(T',F',k')}{dt'dV'd\kb'}\,.
\label{eq:670}
\eeq
The use of the formulas (\ref{eq:610}), (\ref{eq:640}), (\ref{eq:650}) 
allows one to avoid the Lorentz
transformations from the quantities in the c.m. frame of $AA$ collisions
to the ones in the comoving frame of the QGP. It makes the calculations
for an expanding QGP as simple as for a QGP at rest.

Note that from (\ref{eq:610}) it is clear that the $Y$-integration
in (\ref{eq:670}) is dominated by the region $|Y-y|\lsim 1$. Because
the photon emission rate in the QGP rest frame
in the integrand in (\ref{eq:670}) falls rapidly with $k'$, 
and from (\ref{eq:610}) one obtains 
$k'=k\cdot\cosh{(Y-y)}$ (we neglect the
transverse expansion). Since the dominating contribution
in the $\tau$-integration in (\ref{eq:670}) comes from $\tau\lsim 2-3$ fm,
the effective $2$-volume for the integration over $t$ and $z$ is $\sim 5-10$
fm$^2$. It is by a factor of $\sim 10-20$ smaller than that of \cite{T1},
where the $t$- and $z$-integrations have been performed 
for $T=\text{const}$, and $\vb=0$ (which gives $k=k'$) over the region $|z|<t<10$ fm.

\subsection{Model of the fireball}
It is widely believed that the plasma fireball 
is produced in $AA$ collisions  
after  thermalization of the glasma color tubes
created in interaction of the Lorentz-contracted nuclei
\cite{term_gl}.
The typical time of evolution of the glasma color fields
is about several units of $1/Q_s$, where
$Q_s$ ($\sim 1-1.5$ GeV for RHIC and LHC conditions \cite{Lappi_qs}) is 
the saturation scale of the nuclear parton distributions.
It means that even for a very fast thermalization
of the glasma color fields one can apply the formulas obtained
for the equilibrium QGP only at $\tau\gsim 0.2-0.5$ fm.
The thermalization time $\tau\sim 0.2$ fm means practically instantaneous 
process of the glasma thermalization at $\tau\sim 1/Q_s$, and
does not seem to be realistic. Nevertheless, in some analyses
of the photon production \cite{Sinha,Mitra} the authors 
use $\tau_0=0.2$ and $0.1$ fm for RHIC and LHC energies,respectively.
But such small values do not have a theoretical justification.
In the present analysis we use a more realistic value of $\tau_0=0.4$ fm
used in the analysis \cite{Gale_best}. 
To account for qualitatively the fact that 
the process of the QGP production is not instantaneous 
we take the entropy density $\propto \tau$ in the interval $0<\tau<\tau_0$.
However, the contribution of this region is relatively small (due to 
the factor $\tau$  in the integrand in (\ref{eq:670})).

We describe the plasma fireball in the thermalized stage at $\tau>\tau_0$ 
in the Bjorken model \cite{Bjorken} without the transverse expansion
that gives the entropy density $s\propto 1/\tau$. 
For the ideal gas model with $s\propto T^3$ it gives 
$
T=T_0(\tau_0/\tau)^{1/3}
$ in the plasma phase. However, the lattice calculations show \cite{EoS}
that for the temperature range of interest $T\lsim 500$ MeV
the entropy density exhibits a significant deviation from the $s\propto T^3$
dependence. For this reason it seems reasonable \cite{Gale_HG} to determine
the plasma temperature from the temperature dependence of the entropy 
density predicted by lattice calculations. 
In our analysis we determined
$T$ from the entropy density obtained in \cite{EoS}. At $T\sim (1-2)T_c$
it gives the temperature greater than that for the ideal gas dependence
$s\propto T^3$ by $10-20$\%. This relatively small increase in $T$
may be important for the photon emission rate, because its 
$k$-dependence comes mostly from the exponential factor $\exp(-k/T)$
(stemming from the Fermi distribution (\ref{eq:110})),
which at $k\gg T$ is sensitive even to a small variation of $T$.
  
In Bjorken's model the entropy density of the QGP 
at a given impact parameter vector $\bb$ of the $AA$ collision
can be written as
\beq
s(\tau,\ro,Y,\bb)=\frac{1}{\tau}\frac{d S(\ro,Y,\bb)}
{d\ro dY}\,,
\label{eq:680}
\eeq
where ${dS}/{d \ro d Y}$ is the distribution
of the entropy in the impact parameter plane and rapidity.
For simplicity we take a Gaussian distribution of the entropy in
the rapidity
\beq
\frac{dS(\ro,Y,\bb)}{d\ro dY}=\frac{dS(\ro,Y=0,\bb)}{d\ro
  dY}\exp{(-Y^2/2\sigma_Y^2)}\,. 
\label{eq:690}
\eeq
For Au+Au collisions at $\sqrt{s}=0.2$ TeV
we take for the width in $Y$
$\sigma_{Y}=2.63$ which allows to reproduce qualitatively the experimental 
pseudorapidity distribution of the charged particles  $dN_{ch}/d\eta$.
However, the results are not sensitive to
the exact choice of $\sigma_y$, because the dominating contribution
to the $Y$-integral in (\ref{eq:670}) comes from $|Y|\lsim 1$.

We calculate the initial 
density profile in the impact parameter plane of the entropy 
at the proper time $\tau_{0}$
assuming that it is proportional to the charged particle pseudorapidity 
density at $\eta=0$ 
calculated in the two component wounded nucleon 
Glauber model \cite{KN}
\beq
\frac{d N_{ch}(\ro,\bb)}{d\eta d\ro}=
\frac{dN_{ch}^{pp}}{d\eta}
\left[
\frac{(1-\alpha)}{2} \frac{dN_{part}(\ro,\bb)}{d\ro}
+\alpha\frac{dN_{coll}(\ro,\bb)}{d\ro}
\right]\,,
\label{eq:700}
\eeq
where $dN_{ch}^{pp}/d\eta$ is the pseudorapidity multiplicity density
for $pp$ collisions, and 
\beq
\frac{dN_{part}(\ro,\bb)}{d\ro}=
T_A(|\ro-\bb/2|)\left[1-\exp\left(-\sigma_{pp}T_A(|\ro+\bb/2|)\right)\right]
+T_A(|\ro+\bb/2|)
\left[1-\exp\left(-\sigma_{pp}T_A(|\ro-\bb/2|)\right)\right]\,,
\label{eq:710}
\eeq
\beq
\frac{dN_{coll}(\ro,\bb)}{d\ro}=
\sigma_{pp}T_A(|\ro-\bb/2|)T_A(|\ro+\bb/2|)\,.
\label{eq:720}
\eeq
Here $T_A(b)=\int dz n_A(\sqrt{b^2+z^2})$ is the nuclear profile function
calculated with the Woods-Saxon nuclear distribution
\beq
n_{A}(r)=\frac{N}{1+\exp[(r-R_A)/a]}\,,
\label{eq:730}
\eeq
where $N$ is the normalization constant, 
$R_{A}=(1.12A^{1/3}-0.86/A^{1/3})$ fm, $a=0.54$ fm \cite{GLISS2}.
In numerical calculations for Au+Au collisions 
at $\sqrt{s}=0.2$ TeV 
we take $dN_{ch}^{pp}/d\eta=2.65$ and $\sigma_{pp}=35$ mb 
obtained by the UA1 collaboration \cite{UA1_pp}
for non-single diffractive inelastic events. 
We take $\alpha=0.135$ \cite{Z_MCGL}, which allows to describe well the 
data from STAR \cite{STAR1} on the centrality dependence of 
$dN_{ch}/d\eta$ in Au+Au collisions at $\sqrt{s}=0.2$ TeV.
To fix the normalization of the entropy density 
we use the relation 
$
dS/dY{\Big/}dN_{ch}/d\eta\approx 7.67
$ 
obtained in \cite{BM-entropy}.
For central Au+Au collisions 
at $\sqrt{s}=0.2$ TeV this procedure gives the plasma temperature at the center 
of the fireball  $T\approx 465$ MeV  at $\tau=0.4$ fm.
In the space-time integral (\ref{eq:670}) we drop the points which formally 
give $T<T_c$ (we take $T_c=165$ MeV)
at $\tau=\tau_0$. We treat the crossover region at $T\sim T_c$
as a mixed phase assuming that the entropy density in this
phase $\propto 1/\tau$ \cite{Bjorken}, and account for only the QGP phase.
However, the contribution of the space-time region with $T\sim T_c$ 
to the photon spectrum in $AA$ collisions is relatively small even
at $k_T\sim 0.5$ GeV. And at $k_T\gsim 1.5-2$ GeV the contribution of
this space-time region is practically unimportant.

\subsection{Electromagnetic field in the fireball}
For computation of the synchrotron contribution to the photon emission 
rate we need to know the magnitude of the electromagnetic field
in $AA$ collisions in the space-time region occupied by the QGP, 
i.e. even for very optimistic scenarios with
a fast thermalization of the glasma color fields it means the $\tau$-region
$\tau\gsim 0.2$ fm.
Presently, there is no consensus within the heavy ion community 
on the magnitude of the electromagnetic fields in the QGP at
such times.

The magnetic field generated by the Coulomb fields of the colliding nuclei
at $\rb=0$ (the center of the fireball) has
the only nonzero component $B_y$ (for a coordinate frame as shown in Fig.~1). 
At $t=0$ and $\rb=0$ the magnetic field reads \cite{Z_maxw}
\beq
eB_y(t=0,\rb=0)\approx \gamma Z \alpha b/R_A^3\,,
\label{eq:731}
\eeq
and at $t^{2}\gsim (R_{A}^{2}-b^{2}/4)/\gamma^{2}$ 
($b$ is assumed to be $<2R_{A}$)
it is approximately
\beq
eB_{y}(t,\rb=0)\approx \frac{\gamma Z\alpha b}{(b^{2}/4+\gamma^{2}t^{2})^{3/2}}\,.
\label{eq:740}
\eeq
For $t\gg R_{A}/\gamma$    in the region $\rho\ll t\gamma$     
the field has a simple $\rho$-independent form 
\beq
eB_{y}(t,\ro,z=0)\approx
\frac{Z\alpha b}{\gamma^{2}t^{3}}\,.
\label{eq:750}
\eeq
The quantity $R_{A}/\gamma$ is very small: $\sim 0.06$ fm 
for Au+Au collisions at RHIC energy $\sqrt{s}=0.2$ TeV,
and $\sim 0.004$ fm for Pb+Pb collisions at
LHC energy $\sqrt{s}=2.76$ TeV.
For Au+Au collisions at $\sqrt{s}=0.2$ TeV (\ref{eq:750}) gives
\beq
eB_{y}(t,\ro,z=0)\approx m_{\pi}^2 \cdot 10^{-4}
\frac{(b/1\,\text{fm})}{(t/1\,\text{fm})^3}\,.
\label{eq:760}
\eeq
And for Pb+Pb collisions at $\sqrt{s}=2.76$ TeV from  (\ref{eq:750})
we obtain
\beq
eB_{y}(t,\ro,z=0)\approx m_{\pi}^2\cdot 5.5\cdot 10^{-7}
\frac{(b/1\,\text{fm})}{(t/1\,\text{fm})^3}.
\label{eq:770}
\eeq
From these relations we obtain at $t=0.2$ fm 
$eB_y\approx 0.075 m_{\pi}^2$ and $4\cdot 10^{-4} m_{\pi}^2$ for
RHIC and LHC, respectively. Thus, even for very optimistic assumption
on the QGP formation time, the magnitude of the magnetic field
in the initial stage of the QGP phase turns out to be much smaller
than that in the first instant of the $AA$ collision (\ref{eq:731})
($\sim 3m_{\pi}^2(b/R_A)$ and $\sim 40m_{\pi}^2(b/R_A)$ for RHIC 
and LHC, respectively).
From above one sees that from the point of view of the synchrotron 
contribution to the photon emission rate a potentially 
interesting case is Au+Au collisions at RHIC. 
For Pb+Pb collisions at LHC the magnitude of the magnetic field in 
the plasma stage is clearly too small to generate a significant 
synchrotron radiation.
  
The presence of the QGP may modify the electromagnetic fields at later
times due to the conductivity of the QGP.
There was an idea  that the induced currents generated in the
conducting QGP can significantly delay the decay of the magnetic field
\cite{Tuchin_rev}. It is possible if the magnetic lines, at least partly,
are frozen in the QGP similarly to the ordinary conducting materials
\cite{LL8}. However, the analysis performed in \cite{Skokov_B} 
for the QGP with zero velocity has shown that 
for realistic plasma conductivity the effect of the induced
currents is not strong enough to delay considerably the decay of the magnetic 
field. The computations for a realistic expanding plasma fireball
have been performed in \cite{Z_maxw}.
There, by solving Maxwell's equations in the Milne coordinates
$x^{\mu}=(\tau,\rr,Y)$,
it was shown that, formally, at $\tau \gsim 0.5-1$ fm 
the induced currents can generate significant electromagnetic fields
at the center of the fireball 
that are much bigger than
the electromagnetic fields originating from the protons of the colliding 
nuclei. However, for realistic
values of the plasma conductivity, the electromagnetic fields
generated by the induced currents in the fireball turn out to be in
a deep quantum regime when the typical occupation numbers are small. 
In this regime the induced currents lead only 
to a rare emission of single photons 
(with a typical energy about several units of the inverse size of the fireball
(i.e. $\sim 1/R_A$). It is clear that such single-photon processes 
cannot lead to the thermal synchrotron radiation from the QGP. 
In this physical picture of the electromagnetic
response of the QGP we are left only with the synchrotron radiation
related to the electromagnetic field generated by the protons of the colliding
nuclei.
As was shown above at $\tau=0.2$ fm we have 
$eB_y\sim 0.1 m_{\pi}^2$ for Au+Au collisions at $\sqrt{s}\sim 0.2$ TeV, 
To make our estimates of the synchrotron contribution
as optimistic as possible we perform calculations for $eB_y=m_{\pi}^2$.
Note that this value is somewhat larger then the magnitude of magnetic field
obtained in the recent analysis \cite{Tuchin_B}, and than that used in 
calculations of \cite{T1}. 

\begin{figure} [h]
\begin{center}
\epsfig{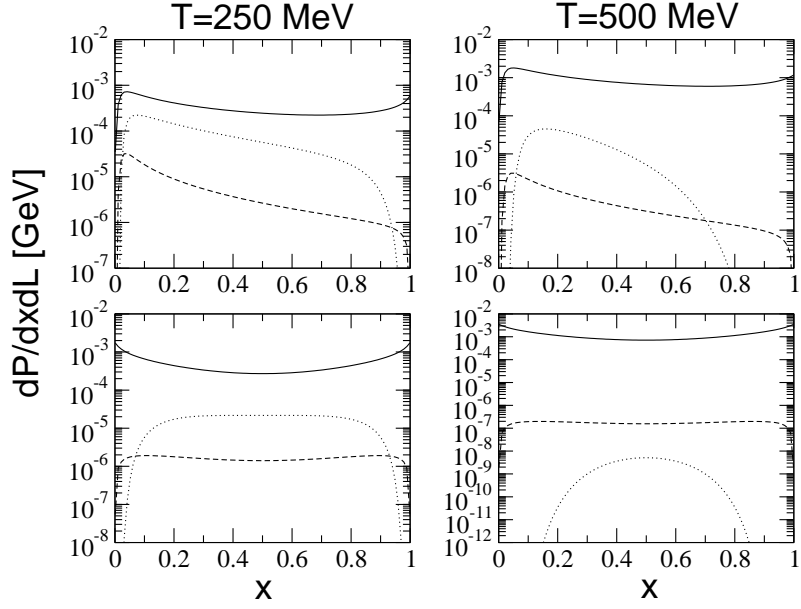}
\end{center}
\caption[.]{The probability 
distribution $dP/dxdL$ for $q\to \gamma q$  
(upper) at $E_q=2$ GeV and $\gamma \to q\bar{q}$ (lower) at $E_{\gamma}=2$ GeV 
at $T=250$ (left) and $500$ (right) 
MeV for $u$ quark. Solid: the contribution of multiple scattering, dotted:
the pure synchrotron contribution, dashed: the synchrotron contribution 
obtained with account for multiple scattering.
The synchrotron contributions are computed for $eB=m_{\pi}^2$.}
\end{figure}
\begin{figure*}[t]
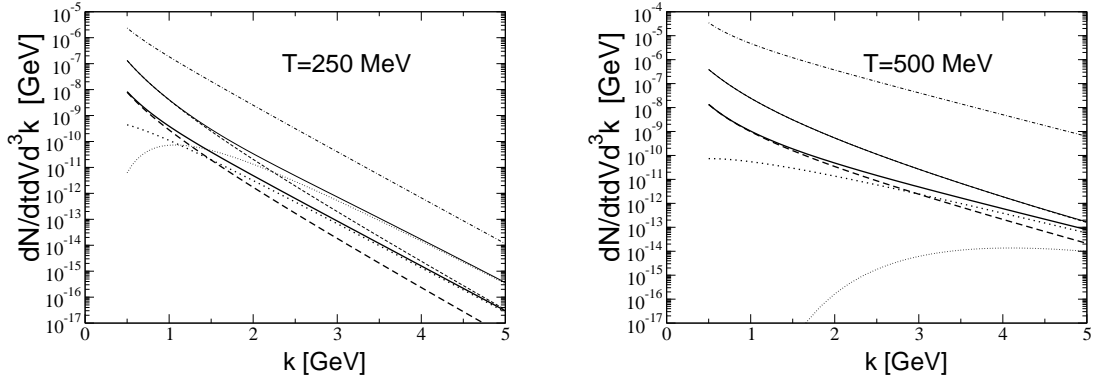

\hspace*{-0.8cm }\epsfig{file=fig3l.eps,height=5cm,clip=,angle=0} 
\hspace*{0.8cm } \epsfig{file=fig3r.eps,height=5cm,clip=,angle=0} 
\begin{minipage}[t]{17.cm}  
\begin{center}
\caption{
The photon emission rate $dN/dtdVd\kb$ in the QGP rest frame 
for $T=250$ (left) and $500$ (right) MeV. Solid: the sum of the 
synchrotron contributions from $q\to \gamma q$ and $q\bar{q}\to \gamma$
processes, dashed:  the synchrotron contribution from $q\to \gamma q$,
dotted: the synchrotron contribution from $q\bar{q}\to \gamma$,
dot-dashed: the sum of the contributions from 
$q\to \gamma q$ and $q\bar{q}\to \gamma$ transitions due to multiple
scattering alone.
The thin solid, dashed, and dotted lines show the predictions 
for the pure synchrotron mechanism,
and the thick ones show the predictions obtained with account for
multiple scattering. 
The synchrotron contributions are computed for $eB=m_{\pi}^2$
(the magnetic field is assumed to be perpendicular to the photon 
momentum).
}
\end{center}
\end{minipage}
\end{figure*}

\section{Numerical results}
In this section we discuss our numerical results
on photon emission from the QGP.
The numerical results are obtained for the quark quasiparticle
mass for $\alpha_s=0.2$. 
The basic ingredients used to 
calculate the photon emission rate from the QGP
are the $x$-spectra of $q\to \gamma q$ and $\gamma \to q\bar{q}$ transitions.
In Fig.~2 we present  the prediction for these spectra for $u$ quark
for $E_{q,\gamma}=2$ GeV
at $T=250$ and $500$ MeV, 
and $eB=m_{\pi}^2$. We show separately 
the contributions
from multiple scattering and the effect of the magnetic field.
For comparison we also show the results
for the purely synchrotron spectrum (i.e., for $\hat{q}=0$).
From Fig.~2 one sees that for $q\to \gamma q$ multiple scattering reduces 
strongly the synchrotron contribution at moderate values of $x$. 
However, even without this suppression
the pure synchrotron contribution is much smaller than
the contribution to the spectrum related to 
multiple scattering of quarks in the QGP.
For the $\gamma\to q\bar{q}$ the pattern of interplay of 
the effects from magnetic field and multiple scattering is more
complicated. At $T=250$ MeV at moderate $x$ the synchrotron contribution 
obtained accounting for multiple scattering is much smaller 
than the one obtained with $\hat{q}=0$. But at $T=500$ MeV
multiple scattering enhances the synchrotron contribution.
However, similarly to the $q\to \gamma q$ process, the synchrotron
contribution turns out to be much smaller than the spectrum 
generated by quark multiple scattering alone.

In Fig.~3 we show the results of the
computation of $dN/dtdVd\kb$
for bremsstrahlung and annihilation and for their sum
at $T=250$ and $500$ MeV.
As in Fig.~2 we present also the curves obtained neglecting
the effect of multiple scattering ($\hat{q}=0$).
One can see that, similarly to Fig.~2,  the contribution
from multiple scattering alone is much bigger than 
the contribution of the synchrotron mechanism. 
The curves for the total synchrotron mechanism 
($q\to \gamma q$ plus $q\bar{q}\to \gamma$)
obtained accounting multiple scattering go considerably below the
ones for the synchrotron contribution for $\hat{q}=0$.
From Fig.~3 one sees that for the synchrotron mechanism
with multiple scattering the contribution
from $q\bar{q}\to \gamma$ process becomes larger than the one
from $q\to \gamma q$ at $k\gsim 1.5$ GeV for $T=250$ MeV and
at $k\gsim 3$ GeV for $T=500$ MeV.
Fig.~3 shows that for a version with multiple scattering
the contribution of the synchrotron mechanism turns out to be practically
negligible as compared to the photon emission due to
ordinary quark multiple scattering in the QGP.

In Figs.~4,~5,~6 we present the results for the photon
spectrum $dN/dy d\kb_T=(1/2\pi k_T)dN/dy dk_T$ 
(averaged over the azimuthal angle) stemming from both
$q\to \gamma q$ and $q\bar{q}\gamma$ processes
for Au+Au collisions at $\sqrt{s}=0.2$ TeV for three centrality bins
$0-20$\%, $20-40$\%, and $40-60$\%.
The theoretical curves have been obtained integrating in (\ref{eq:670}) 
up to 
$\tau_{max}=10$ fm. The calculations with $\tau_{max}=R_A\approx 6.4$ fm 
give very similar results at $k_T\gsim 1.5$ GeV,
and at $k_T\sim 0.5-0.75$ GeV the photon spectrum is reduced 
by $\sim 30-40$\%.    
At $k_T\gsim 1.5$ GeV the results are only weakly sensitive 
to $\tau_{max}$, because the main contribution at $k_T\gg T_0$ comes 
from the hottest space-time region of the QGP with $\tau$ up to
several units of $\tau_0$.
As in Fig.~3 we show the results for multiple scattering
alone and for the two versions of the synchrotron contribution.   
One sees that multiple scattering reduces strongly 
the synchrotron contribution. It is important that for both the versions
of the synchrotron contribution the effect is much smaller (by a factor 
of $\sim 10^3-10^4$) than the contribution from multiple scattering.
Our calculations show that the azimuthal asymmetry $v_2$ 
for the synchrotron contribution alone is large ($\sim 0.5$).
However, since the relative contribution of the synchrotron 
mechanism to the photon emission rate 
is very small, its effect on the observable $v_2$ turns out to
be negligible as well.
We also present in Figs.~4,~5,~6 the sum of our contribution from multiple
scattering and the  
the LO contribution from $2\to 2$ processes 
$q(\bar{q})g\to \gamma q(\bar{q})$ 
and $q\bar{q}\to \gamma g$ 
in the form obtained in \cite{AMY1}.
Although a detailed analysis of the experimental data on the direct photons
in $AA$ collisions is not a purpose of this paper,
in Figs.~4,~5,~6 we also plot the data from PHENIX 
\cite{PHENIX_ph_PR} obtained after subtraction of the $N_{coll}$ scaled
photon spectrum for $pp$ collisions.
One can see that the theoretical curves
for the sum of the contribution from the collinear processes
$q\to \gamma q$ and $q\bar{q}\gamma$ and the LO mechanisms
underestimate the data by a factor of $\sim 2-4$. It is slightly 
bigger than found in the analysis \cite{Gale_best} ($\sim 1.5-3.5$). However, in 
\cite{Gale_best}, in addition to the photon emission from the QGP,
the radiation from the hadron gas has been included, which is neglected
in our calculations. 
\begin{figure} 
\begin{center}
\epsfig{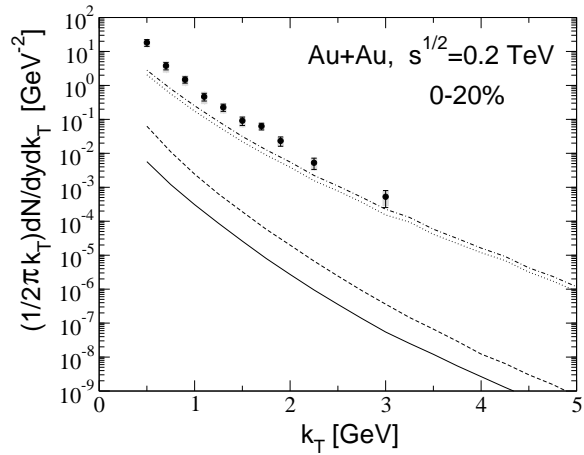}
\end{center}
\caption[.]{
The photon spectrum  $(1/2\pi k_T)dN/dydk_T$ 
averaged over the azimuthal angle for Au+Au collisions at 
$\sqrt{s}=0.2$ TeV in the $0-20$\% centrality range.
Solid: the sum of the 
synchrotron contributions from $q\to \gamma q$ and $q\bar{q}\to \gamma$
processes calculated with account for multiple scattering, 
dashed:  the same as solid but without the effect of 
multiple scattering, dotted: the contribution
from $q\to \gamma q$ and $q\bar{q}\to \gamma$
processes due to quark multiple scattering alone, 
dot-dashed: the sum  of 
the contributions from $q\to \gamma q$ and $q\bar{q}\to \gamma$
processes due to quark multiple scattering and
the contribution of the LO $2\to 2$ processes in the form obtained
in \cite{AMY1}. The data are from Ref.~\cite{PHENIX_ph_PR}.
}
\end{figure}
\begin{figure} [h]
\begin{center}
\epsfig{file=fig5.eps,height=6cm,angle=0}
\end{center}
\caption[.]{
Same as in Fig.~4 for $20-40$\% centrality bin.}
\end{figure}
\begin{figure} [h]
\begin{center}
\epsfig{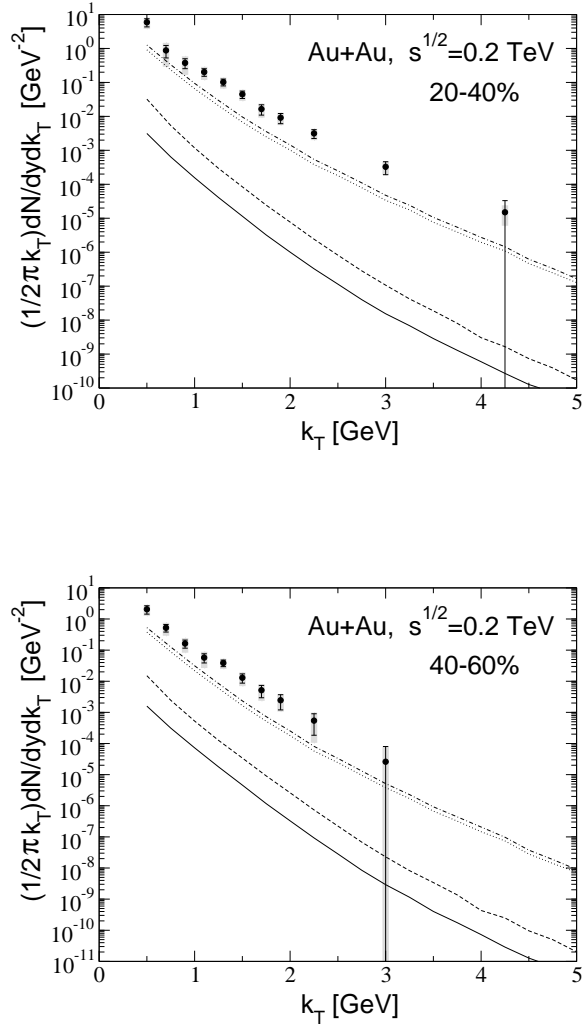}
\end{center}
\caption[.]{
Same as in Fig.~4 for $40-60$\% centrality bin.}
\end{figure}

Thus, our calculations show that even for clearly too optimistic 
value of the magnetic field the effect of the synchrotron mechanism is 
very small. 
For more realistic field $eB\sim 0.1 m_{\pi}^2$
the synchrotron contribution is smaller by a factor of $\sim 10^2$.
It leads to the conclusion that 
for RHIC and LHC conditions the synchrotron mechanism
cannot be important neither for the azimuthally averaged photon spectrum
nor for the azimuthal asymmetry $v_2$\footnote{ 
Assuming that our collinear 
formulas are qualitatively valid at $L_f/R_L\sim 1-3$ we have found that 
to obtain the photon $v_2$ with a magnitude comparable to the measured 
$v_2$ at $k_T\sim 1-3$ GeV  one should assume that $eB\sim (30-70) m_{\pi}^2$.
However, such strong fields in the QGP stage are clearly unrealistic because
they are by a factor of $\sim 10-20$ bigger than even the magnetic field in 
the first instant after the $AA$ collision 
(we consider Au+Au collisions at $\sqrt{s}=0.2$ TeV, and 
take $t\lsim R_A/\gamma$ and $b\sim R_A$).
Also, the electromagnetic energy density for such fields turns out to 
be too large (of the order of the thermal plasma energy 
at $T\sim 600$ MeV) both for the RHIC and LHC conditions. 
These arguments show that the scenario with 
$eB\gg m_{\pi}^2$, which could formally give a reasonable agreement
with experimental data on $v_2$, can be rejected. 
}.

Our results are in strong disagreement with the recent analysis
\cite{T1}, where a rather large effect of magnetic
field was found. 
At $k_T\sim 1-3$ GeV our synchrotron contribution obtained without 
the effect of multiple scattering shown in Figs.~4,~5,~6 
by a factor of $\sim 10^2-10^3$ smaller than that from \cite{T1}.
In \cite{T1} the $k_T$ photon spectrum was calculated
for the QGP at rest and $T=\text{const}$.
As noted in the Introduction and in Sec.~3,
this approximation should overestimate the photon emission 
rate (at least by a factor of $\sim 10$). However, 
the major source of the difference between our
results and that of \cite{T1} is probably the different 
choice of the quark masses. We use
for the quark mass the quark quasiparticle mass, while
in \cite{T1} the current quark masses have been used. 
The theoretical basis for the use of the quasiparticle quark mass is
same as in the AMY scheme \cite{AMY1}, where quarks acquire
a dynamical thermal mass $\sim gT$ after the HTL resummation.
As was demonstrated in Sec.~2 the adding of the external magnetic field
does not change the physical picture of the collinear photon emission.
We checked that for the photon momentum $k\sim 1-3$ GeV and $T\sim 250-500$
MeV the replacement of the thermal quark mass by the current one
increases the pure synchrotron contribution by a factor of $\sim 10-200$.
Note that for the synchrotron contribution obtained accounting for
multiple scattering, i.e. for nonzero $\hat{q}$, the replacement of the thermal
quark mass by the current one gives a relatively small enhancement ($\lsim
1.5$). It is connected with the fact that 
the coherence length of the photon/gluon emission 
in the presence of multiple scattering
remains finite even for massless partons\footnote{In terms of Eq. (\ref{eq:50})
it means that the quantity $S_{LPM}/m_q^2$ is finite in the limit
$m_q\to 0$.}. Note that just for this reason
the parton energy loss is well defined in the massless limit \cite{BDMPS}.
It is worth noting that the fact that the synchrotron contribution 
in the presence of multiple
scattering remains small even for massless quarks shows that 
it should be small also for the scenario of a strongly coupled QGP 
with a very small thermal quark mass \cite{mq_sQGP}.

\section{Summary}
We have developed a formalism for evaluation of the
photon emission from the QGP with external electromagnetic field 
due to the collinear processes
$q\to \gamma q$ and $q\bar{q}\to \gamma$.
Within this formalism we have studied the effect of magnetic field
on the photon emission rate from the QGP in $AA$ collisions
for a realistic model of the plasma fireball.
We showed that that multiple scattering reduces considerably the effect
of magnetic field.
We found that even for an extremely optimistic 
assumption on the magnitude of magnetic field
($eB\sim m_{\pi}^2$)
the effect of magnetic field on the photon emission
in $AA$ collisions is very small. For more realistic
fields ($eB\sim 0.1 m_{\pi}^2$) the effect is practically negligible.
For this reason, we conclude that the synchrotron mechanism cannot 
lead to a considerable azimuthal asymmetry in the photon emission rate
in $AA$ collisions.
Our calculations show that due to multiple
scattering the synchrotron contribution is small even 
for massless quarks. For this reason 
for the scenario of a strongly coupled QGP with
a very small thermal quark mass \cite{mq_sQGP}
the effect of magnetic field on the photon emission 
should remain small.

\begin{acknowledgments} 	
I thank P. Aurenche for useful discussions in the initial stage
of this work. I am grateful to K. Tuchin for informing 
me about the values of parameters used in the numerical
calculations of \cite{T1}.  
This work has been supported by the RScF grant 16-12-10151.
\end{acknowledgments}

\end{document}